\documentclass[aps,prl,twocolumn,superscriptaddress,longbibliography,nofootinbib]{revtex4-2}
\usepackage{amsfonts}
\usepackage{mathrsfs,soul}
\usepackage{amsmath}
\usepackage{color}
\usepackage{graphicx}
\usepackage{bm}
\usepackage{amssymb}
\usepackage{xspace}
\usepackage{epstopdf}
\usepackage{dcolumn}
\usepackage{longtable}
\usepackage{multirow}
\usepackage{float}
\usepackage{comment}
\usepackage[bottom]{footmisc}

\usepackage[colorlinks=true, letterpaper=true, pdfstartview=FitV,  linkcolor=blue, citecolor=blue, urlcolor=blue]{hyperref}

\begin{document}

\title{Quantized Spin Hall Effect in Three-Dimensional Nodal-Ring Semimetal: Geometric Scaling and Symmetry‑Engineered Spin Response}
\author{Jiali Chen}
\affiliation{Key Lab of Advanced Optoelectronic Quantum Architecture and Measurement (MOE), Beijing Key Laboratory of Quantum Matter State Control and Ultra-Precision Measurement Technology, and School of Physics, Beijing Institute of Technology, Beijing 100081, China}
\author{Chaoxi Cui}
\affiliation{Key Lab of Advanced Optoelectronic Quantum Architecture and Measurement (MOE), Beijing Key Laboratory of Quantum Matter State Control and Ultra-Precision Measurement Technology, and School of Physics, Beijing Institute of Technology, Beijing 100081, China}
\author{Zhi-Ming Yu}
\affiliation{Key Lab of Advanced Optoelectronic Quantum Architecture and Measurement (MOE), Beijing Key Laboratory of Quantum Matter State Control and Ultra-Precision Measurement Technology, and School of Physics, Beijing Institute of Technology, Beijing 100081, China}
\affiliation{International Center for Quantum Materials, Beijing Institute of Technology, Zhuhai 519000, China}
\author{Wei Jiang}
\email{wjiang@bit.edu.cn}
\affiliation{Key Lab of Advanced Optoelectronic Quantum Architecture and Measurement (MOE), Beijing Key Laboratory of Quantum Matter State Control and Ultra-Precision Measurement Technology, and School of Physics, Beijing Institute of Technology, Beijing 100081, China}
\affiliation{International Center for Quantum Materials, Beijing Institute of Technology, Zhuhai 519000, China}
\author{Yugui Yao}
\email{ygyao@bit.edu.cn}
\affiliation{Key Lab of Advanced Optoelectronic Quantum Architecture and Measurement (MOE), Beijing Key Laboratory of Quantum Matter State Control and Ultra-Precision Measurement Technology, and School of Physics, Beijing Institute of Technology, Beijing 100081, China}
\affiliation{International Center for Quantum Materials, Beijing Institute of Technology, Zhuhai 519000, China}


\begin{abstract}

The anomalous Hall conductivity in magnetic Weyl semimetals scales linearly with the momentum separation between Weyl nodes, establishing a geometric paradigm for three-dimensional Hall responses. Here we discover an analogous phenomenon in the spin Hall effect: a quantized spin Hall conductivity (SHC) in nodal‑ring semimetals that scales linearly with the nodal‑ring radius $R$. From an ideal model with a single nodal ring, we derive analytically that the SHC inside the spin-orbit-coupled gap obeys $\sigma_{\alpha \beta}^{S, 3D}=\sigma_0^{S,2D} \cdot (\pi R/2 \pi)$, where $\sigma_0^{S,2D}=(e^2/h) \cdot (\hbar/2 e)$ is the two-dimensional quantum spin Hall conductance. Crucially, the symmetry of the spin-orbit coupling acts as an independent switch: Rashba coupling generates purely conventional SHC components, while Weyl coupling additionally activates unconventional ones, providing separate control over response magnitude and tensor symmetry. We validate this principle in yttrium nitride, where strain tunes $R$ and symmetry breaking toggles between response types. Our work establishes a new paradigm for engineering quantized geometric responses in three dimensions, opening pathways to tailored spin–orbit functionalities.

\end{abstract}

\maketitle


\paragraph{Introduction.} Topological quantum states have fundamentally reshaped condensed matter physics \cite{kane2TopologicalOrder2005,hasanColloquiumTopologicalInsulators2010,qiTopologicalInsulatorsSuperconductors2011a,Hasan2015TIchapter}, with quantized transport phenomena in two-dimensional (2D) systems standing as a defining achievement. The quantum Hall \cite{klitzingNewMethodHighAccuracy1980,thoulessQuantizedHallConductance1982}, quantum anomalous Hall (QAH)~\cite{nagaosaAnomalousHallEffect2010,jinThreedimensionalQuantumAnomalous2018}, and quantum spin Hall (QSH) effects \cite{kaneQuantumSpinHall2005,doi:10.1143/JPSJ.77.031007,bernevigQuantumSpinHall2006a,shengQuantumSpinHallEffect2006,dengTwistedNodalWires2022,PhysRevB.91.081111} each originate from one-dimensional chiral or helical edge channels whose conductance is precisely quantized in units of topological invariants~\cite{checkelskyTrajectoryAnomalousHall2014, chiuClassificationTopologicalQuantum2016, bansilColloquiumTopologicalBand2016}. Extending such quantization to three dimensions (3D) represents a fundamental open frontier \cite{PhysRevB.45.13488,PhysRevLett.86.1062,PhysRevLett.99.146804,zhangThreedimensionalQuantumAnomalous2025,royTopologicalPhasesQuantum2009,zhao3DQuantumHall2023}, which is challenging due to the 2D nature of topological surface states \cite{qiQuantumSpinHall2010,zhangTopologicalInsulatorsBi2Se32009}. Recent breakthroughs have established that topological semimetals can host geometric scaling laws that play a role in 3D analogous to topological invariants in 2D, offering practical knobs for performance optimization even in the absence of strict quantization~\cite{burkovAnomalousHallEffect2014, Jiang2021PRL,yangQuantumHallEffects2011}. 

\begin{figure}[h]
    \centering
    \includegraphics[width=0.47\textwidth]{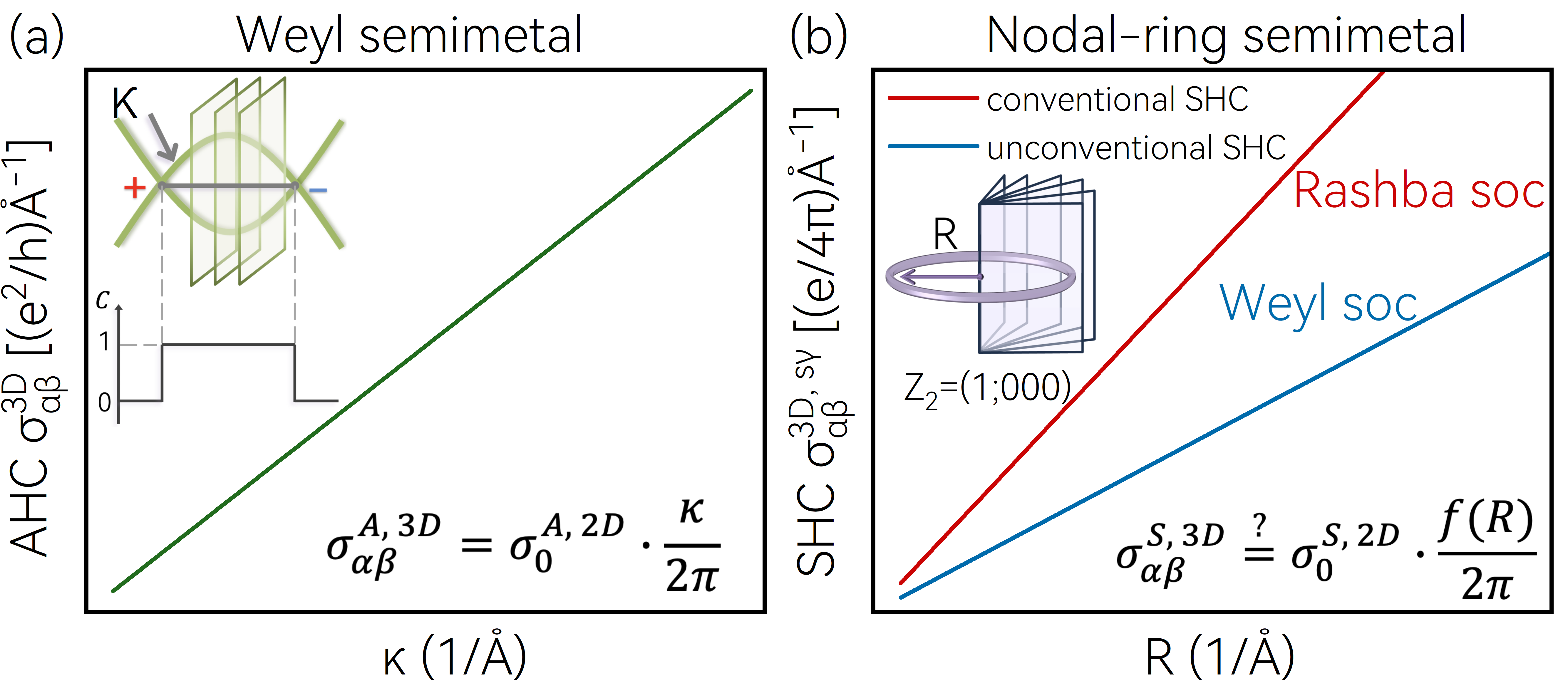}
    \caption{Geometric scaling of Hall responses in topological semimetals.
(a) In a magnetic Weyl semimetal, the AHC ($\sigma^{A,3\mathrm{D}}_{\alpha\beta}$) scales with the Weyl-pair momentum separation $\kappa$. The inset shows the Chern number $C$ associated with 2D $k$-slices.
(b) In a SOC-gapped nodal-ring semimetal with $Z_{2}=(1;000)$, the SHC scales with the nodal-ring radius $R$ as $\sigma_{\alpha\beta}^{S,3D} \propto f(R)$. $\sigma^{A,2\mathrm{D}}_{0}=e^{2}/h$ and $\sigma^{S,2\mathrm{D}}_{0}=e/4 \pi$ are the quantized anomalous Hall and spin Hall conductances, respectively \cite{kaneQuantumSpinHall2005}. The SOC form (Rashba versus Weyl) acts as a symmetry switch that tunes the relative weights of conventional and unconventional SHC components.}
\label{p0} 
\end{figure}

A paradigmatic example is the anomalous Hall conductivity (AHC) \cite{nagaosaAnomalousHallEffect2010} in magnetic Weyl semimetals \cite{burkovAnomalousHallEffect2014, Jiang2021PRL,yangQuantumHallEffects2011}, which scales linearly with the momentum separation between Weyl pairs $\kappa$, a continuous geometric parameter rather than a quantized topological invariant [Fig.~\ref{p0}(a)]~\cite{kimThreedimensionalQuantumAnomalous2018}. This raises a natural question: can the spin Hall conductivity~\cite{sinovaUniversalIntrinsicSpin2004,murakamiSU2NonAbelian2004,yangClassificationStableThreedimensional2014,wanTopologicalSemimetalFermiarc2011,burkovTopologicalNodalSemimetals2011,schoopDiracConeProtected2016,sunStrongIntrinsicSpin2016} also exhibit a quantized geometric scaling, and if so, what physical parameter would govern it? Nodal-ring semimetals~\cite{chiuClassificationReflectionsymmetryprotectedTopological2014}, with their extended band degeneracies, can generate strong spin Berry curvature (SBC) and thus large SHC~\cite{zhangDifferentTypesSpin2021} when gapped by spin-orbit coupling (SOC), offering an ideal platform to address this question. However, the connection between SHC and nodal‑ring topology remains largely empirical~\cite{zhangDifferentTypesSpin2021,Chenintrinsic2024,Dehghanenhancing2025}, and a universal scaling law or quantized behavior analogous to the Weyl‑semimetal AHC has not been established.

Crucially, different components of the SHC tensor offer complementary spintronic functions \cite{manchonCurrentinducedSpinorbitTorques2019}. Conventional components, with mutually orthogonal charge-current, spin-current, and spin-polarization directions, generate in-plane spin currents ideally suited for spin-orbit torque magnetic random-access memory, where decoupled read and write paths enable high endurance \cite{sinovaSpinHallEffects2015}. Unconventional components instead provide out-of-plane spin currents for field-free switching of perpendicular magnetic memory and programmable spin logic \cite{PhysRevB.92.041101,zhangDifferentTypesSpin2021,zhuSwitchingPerpendicularMagnetization2023,shaoRoadmapSpinOrbit2021}. A central challenge in the field has been to achieve independent control over these distinct SHC components within a single material system, particularly the unconventional ones \cite{songCoexistenceLargeConventional2020,macneillControlSpinOrbit2017,liuSymmetrydependentFieldfreeSwitching2021}. It therefore remains an open and intriguing question whether different forms of SOC can selectively activate targeted SHC components and independently tune their magnitudes for application-specific devices [Fig.~\ref{p0}(b)].

In this letter, we establish a universal quantized geometric scaling law linking the SHC in 3D nodal-ring semimetals to the nodal-ring radius $R$, with a proportionality constant of $(e^2/h) \cdot (\hbar/2 e)=e/4 \pi$. We demonstrate that the SOC symmetry dictates the SHC tensor structure: Rashba- and Weyl-type SOC generate purely conventional and unconventional components, respectively. We validate this scaling in realistic materials such as yttrium nitride (YN), where strain tunes $R$ and the SHC magnitude, while symmetry breaking tailors the SOC type. These results enable independent and precise control of the amplitude and tensor symmetry of the spin Hall response through strain and SOC engineering.

\begin{figure}[h]
    \centering
    \includegraphics[width=0.48\textwidth]{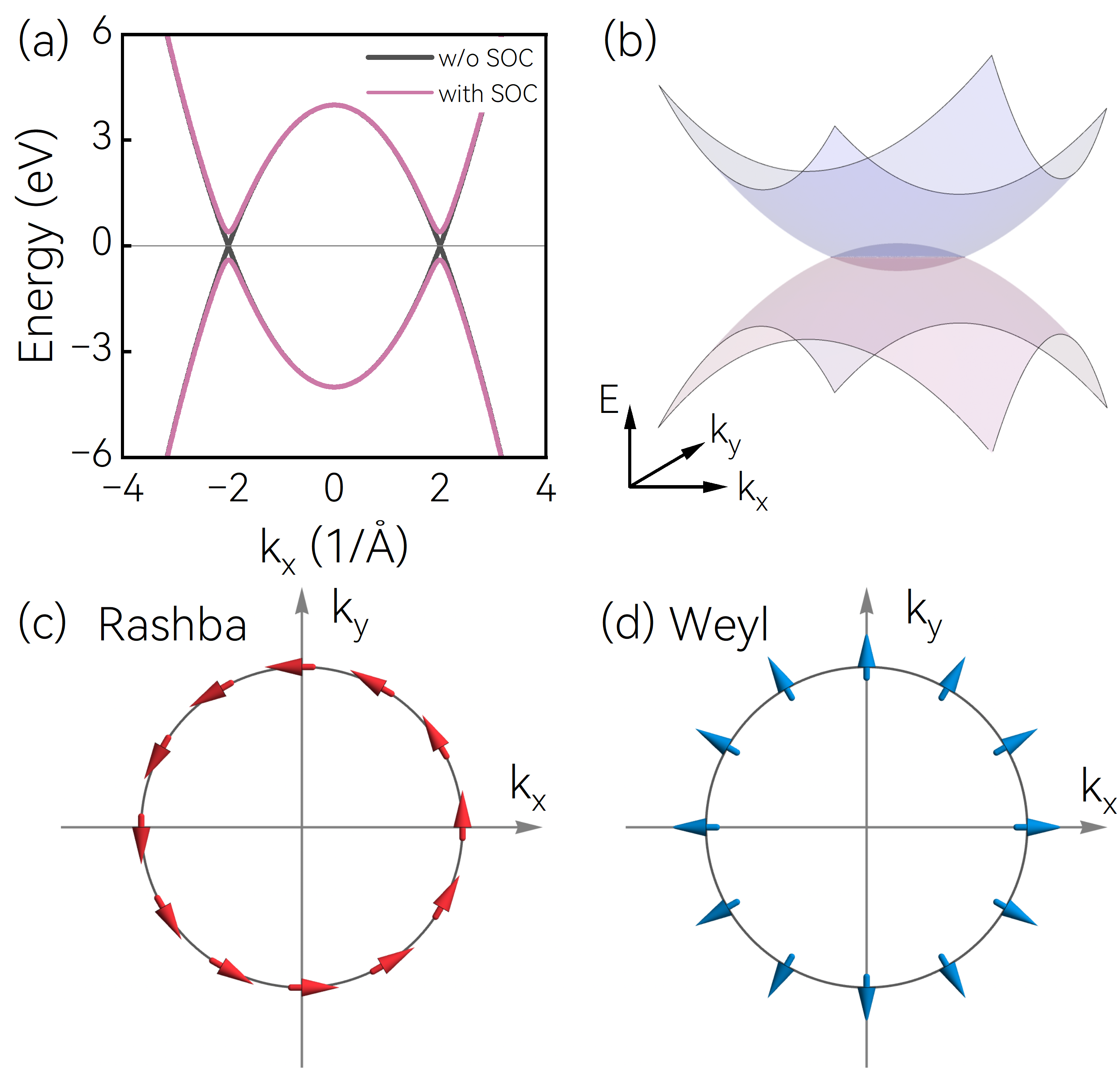}
    \caption{Band structure and generic spin texture of the nodal ring with \(D_0 = 1\,\mathrm{eV}\, \text{\AA}^2,\, v_z = 2\,\mathrm{eV}\, \text{\AA},\, R = 2\,\text{\AA}^{-1},\, \lambda_{\text{so}} = 0.2\,\mathrm{eV}\, \text{\AA}\). (a) Band dispersion of the model Hamiltonian without (black line) and with SOC (pink line) along \(k_x\) with fixed \(k_y = k_z = 0\). (b) 3D band structure of \(H_0\) showing the circular nodal ring at \(k_z = 0\). (c,d) Surface-state spin textures for Rashba- and Weyl-type SOC, respectively.}\label{p1} 
\end{figure}

\paragraph{Nodal-ring $k\cdot p$ toy model.} To uncover the geometric origin of the quantized SHC, we construct a simple 3D model with a single nodal ring at the Fermi level. 
In an orbital basis with in-plane rotational symmetry (e.g., $d_{z^2}$ and $p_z$), the low-energy Hamiltonian near the $\Gamma$ point is:
\begin{equation}
H_0 (\mathbf{k}) = D_0 (k_x^2 + k_y^2 - R^2)\,\tau_z + v_z k_z\,\tau_y,\label{e1}
\end{equation}
where $\tau_i$ (\(i = x, y, z\)) are Pauli matrices acting on the orbital subspace, and $D_0$, $v_z$, and $R$ denote the in-plane curvature of the mass term, out-of-plane velocity, and nodal-ring radius in the $k_z=0$ plane. The continuous rotational symmetry $C_{\infty[001]}$ in the $k_x-k_y$ plane and mirror symmetries $M_{110}$ and $M_{001}$ keep the nodal ring perfectly circular. For illustration we take $D_0 = 1\,\mathrm{eV}\,\text{\AA}^2$, $v_z = 2\,\mathrm{eV}\,\text{\AA}$, and $R = 2\,\text{\AA}^{-1}$, as shown in Figs.~\ref{p1}(a,b).

SOC is required to generate a nonzero SBC and thus an intrinsic SHC \cite{murakamiSpinHallInsulator2004}. To understand how different SOC geometries affect the SHC, we introduce two representative forms that respect the symmetries of $H_0$: a Rashba‑type term coupling spin perpendicular to the in‑plane momentum \cite{manchonNewPerspectivesRashba2015}, and a Weyl‑type term coupling spin radially \cite{wanTopologicalSemimetalFermiarc2011}. In the spin subspace (described by Pauli matrices $\sigma_i$) they read
\begin{align}
H_\mathrm{R} (\mathbf{k})&= -k_y\,\sigma_x + k_x\,\sigma_y, \\
H_\mathrm{W} (\mathbf{k})&= k_x\,\sigma_x + k_y\,\sigma_y.
\end{align}
To induce a nontrivial spin texture and open a global gap, we use an off-diagonal orbital coupling, such as $\tau_x$ or $\tau_y$, which hybridizes the two crossing orbital sectors and opens a global SOC gap along the entire ring [Fig.~\ref{p1}(a), pink line], thereby driving the
system into a strong topological-insulator phase. The full SOC Hamiltonian is then $H^{\mathrm{R/W}}_\mathrm{soc}(\mathbf{k}) =\lambda_{\text{so}} H_{\mathrm{R/W}} (\mathbf{k})\otimes \tau_x$, with $\lambda_{\text{so}}$ the SOC strength. The resulting surface states within the gap inherit the characteristic spin texture: tangential for Rashba SOC [Fig.~\ref{p1}(c)] and radial for Weyl SOC [Fig.~\ref{p1}(d)]. Though SOC lifts the nodal-ring degeneracy, \(R\) remains a well-defined geometric parameter inherited from the parent gapless nodal ring; further details are provided in the Supplemental Material~\cite{SM}.
\nocite{blochlProjectorAugmentedwaveMethod1994, gaoIrvspObtainIrreducible2021, grothKwantSoftwarePackage2014, guoIntrinsicSpinHall2008, liuUnconventionalSpinTextures2024, marzariMaximallyLocalizedGeneralized1997a, perdewGeneralizedGradientApproximation1996, PhysRevB.59.1758, PhysRevB.89.155114, PhysRevB.93.205104, souzaMaximallyLocalizedWannier2001a, wangVASPKITUserfriendlyInterface2021, wuWannierToolsOpensourceSoftware2018, zakBerrysPhaseEnergy1989}

\paragraph{SHC quantization.} The minimal model described above is analytically tractable and captures the essential physics of the nodal ring. We compute the intrinsic SHC using the clean-limit Kubo formula \cite{sinovaUniversalIntrinsicSpin2004,guoInitioCalculationIntrinsic2005,xiaoBerryPhaseEffects2010,vanderbiltBerryPhasesElectronic2018,gradhandFirstprincipleCalculationsBerry2012}
\begin{equation}
\sigma_{\alpha \beta}^{\operatorname{s} \gamma}
= e \hbar \int_{\mathrm{BZ}} \frac{d^3\mathbf{k}}{(2 \pi)^3}
\sum_n f_{n\mathbf{k}}\, \Omega_{n, \alpha \beta}^{\operatorname{s} \gamma}(\mathbf{k}),
\label{ek}
\end{equation}
where the integral is over the Brillouin zone (BZ), and the SBC $\Omega_{n, \alpha \beta}^{\operatorname{s} \gamma}(\mathbf{k})$ is a $k$-resolved band-projected term, analogous to the ordinary Berry curvature \cite{yaoFirstPrinciplesCalculation2004}:  
\begin{equation}\textstyle
\Omega_{n, \alpha \beta}^{\operatorname{s} \gamma}(\mathbf{k})= \sum_{m \neq n} \frac{-2 \operatorname{Im}\left[\left\langle n \mathbf{k}\left|\hat{\jmath}_\alpha^{s, \gamma}\right| m \mathbf{k}\right\rangle\left\langle m \mathbf{k}\left|\hat{v}_\beta\right| n \mathbf{k}\right\rangle\right]}{\left(E_{n \mathbf{k}}-E_{m \mathbf{k}}\right)^2}.\label{ek1} 
\end{equation} 
Here, $\alpha$, $\beta$, and $\gamma$ denote the spin-current, electric-field, and spin-polarization directions, respectively. The spin-current operator is $\hat{\jmath}_\alpha^{s, \gamma}=\frac{1}{2}\left\{\hat{v}_\alpha, \hat{s}_\gamma\right\}$, with the spin operator $\hat{s}_\gamma=\frac{\hbar}{2} \hat{\sigma}_\gamma$ and the velocity operator $\hat{v}_\beta=\frac{1}{\hbar}\frac{\partial H}{\partial k_\beta}$.

In three dimensions the SHC is a third‑rank tensor with 27 components, whose nonzero components are determined by the magnetic point‑group symmetry of the system~\cite{seemannSymmetryimposedShapeLinear2015,gallegoAutomaticCalculationSymmetryadapted2019,kleinerSpaceTimeSymmetryTransport1966}. Following the standard classification \cite{songCoexistenceLargeConventional2020,zhangDifferentTypesSpin2021}, a component is called \emph{conventional} when the spin polarization $\gamma$, spin-current direction $\alpha$, and electric-field direction $\beta$ are mutually orthogonal (e.g., $\sigma^{x}_{yz},\sigma^{y}_{zx},\sigma^{z}_{xy}$). Components that do not satisfy this orthogonality condition are termed \emph{unconventional} and are usually forbidden by mirror or twofold rotational symmetries.

\begin{figure}[ht]
    \centering
    \includegraphics[width=0.48\textwidth]{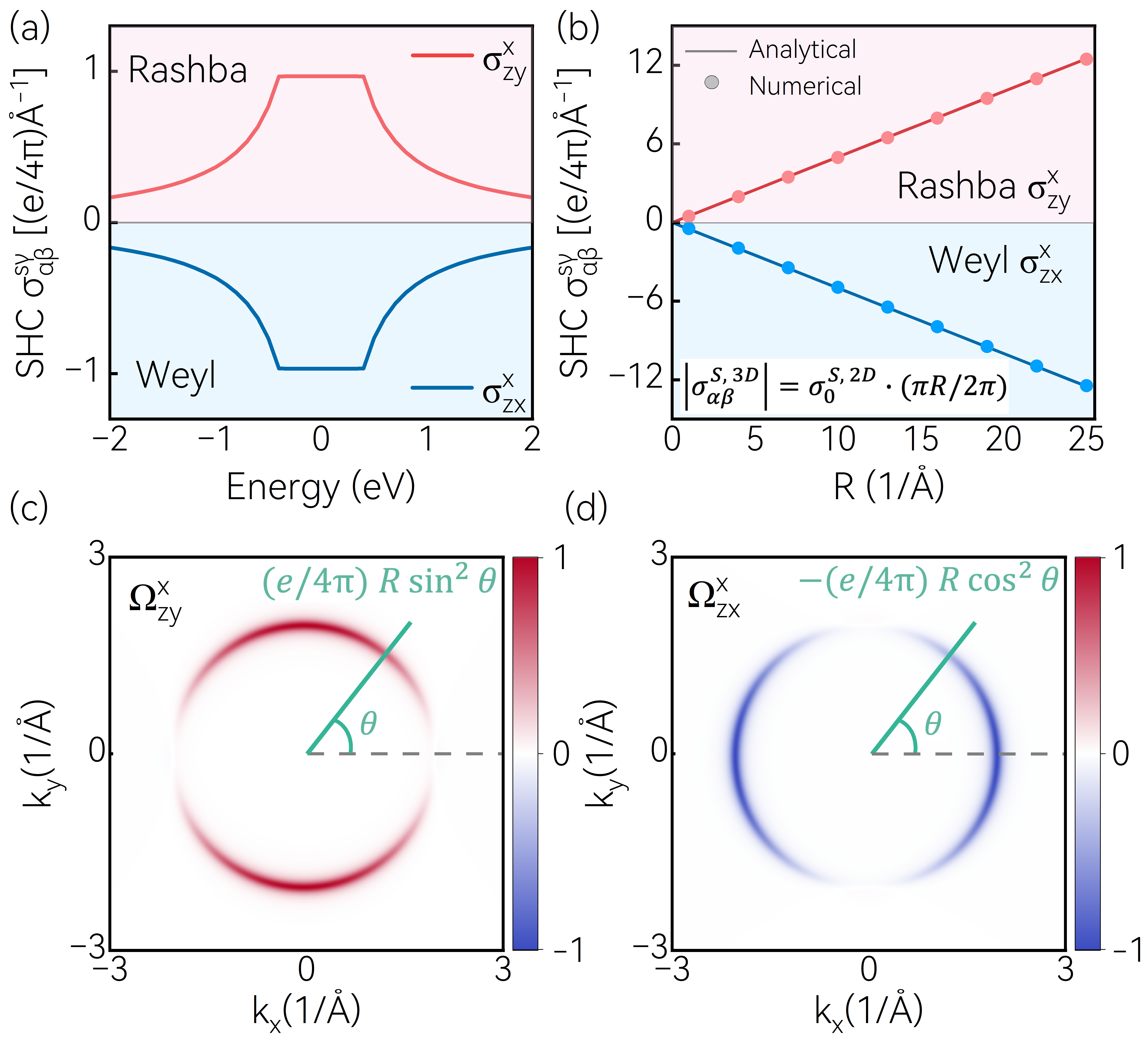}
    \caption{Calculated SHC and SBC distributions. (a) Energy-dependent SHC components for Rashba (red) and Weyl (blue) SOC. (b) SHC versus nodal-ring radius $R$; solid lines are analytical Kubo results and symbols denote numerical calculations.
    (c,d) SBC distribution on the $k_z = 0$ plane for Rashba ($\Omega_{zy}^{\operatorname{spin} x}$) and Weyl ($\Omega_{zx}^{\operatorname{spin} x}$) cases, respectively; the color bar is in arbitrary units. The green line indicates a fixed-\(\theta\) cut along which the SBC is integrated over \((r,k_z)\). Unless otherwise specified (e.g., varying $R$ in panel b), the parameters are
    \(D_0 = 1\,\mathrm{eV}\, \text{\AA}^2,\, v_z = 2\,\mathrm{eV}\, \text{\AA},\, R = 2\,\text{\AA}^{-1},\, \lambda_{\text{so}} = 0.2\,\mathrm{eV}\, \text{\AA}\).
    }\label{p2} 
\end{figure}

SOC symmetry selects the allowed SHC tensor components. Since the Rashba-type coupling $H_\mathrm{R} \otimes \tau_x$ preserves all mirrors of $H_0$, only the six conventional components survive. In contrast, the Weyl-type coupling $H_\mathrm{W} \otimes \tau_x$ breaks the mirror $M_{110}$, thereby allowing seven additional unconventional components ($\sigma^{x}_{xz}, \sigma^{x}_{zx}, \sigma^{y}_{yz}, \sigma^{y}_{zy}, \sigma^{z}_{xx}, \sigma^{z}_{yy}$, and $\sigma^{z}_{zz}$). In both cases, the remaining $C_{\infty[001]}$ symmetry enforces in-plane isotropy, leading to equal magnitudes of the in-plane components, i.e.,  $\sigma^{x}_{yz} = -\sigma^{y}_{xz}$, while the out-of-plane-spin-polarized components vanish in the present model. Numerical Kubo calculations confirm this symmetry analysis, as summarized by the full Fermi-level SHC tensors in Table~IV of \cite{SM}. Importantly, the energy-dependent SHC shows clear plateaus inside the SOC‑induced bulk gap [Fig.~\ref{p2}(a)], signaling a quantized spin Hall response.

To understand the SHC scaling, we systematically vary the SOC strength $\lambda_{\mathrm{so}}$ and the nodal-ring dispersion scale $RD_0$. Contrary to the common intuition that stronger SOC always enhances the SHC \cite{sinovaSpinHallEffects2015}, we find a competition: larger $RD_0$ sharpens the SBC distribution around the gapped ring and larger $R$ expands its momentum-space support, whereas increasing $\lambda_{\mathrm{so}}$ enlarges the gap and suppresses the SHC through the Kubo denominator. This interplay is captured by the dimensionless parameter $A=\lambda_{\mathrm{so}}/(RD_0)$. In the perturbative regime $A\lesssim0.1$, where the
SOC-induced gap is small compared with the nodal-ring bandwidth, a Taylor expansion in $A$ yields analytical expressions for the nonzero SHC components \cite{SM}. Remarkably, each component exhibits a quantized linear dependence on the nodal‑ring radius $R$, 
\begin{equation}
\sigma_{\alpha \beta}^{S, 3D} = \sigma_{0}^{S, 2D} \cdot \frac{\pi R}{2\pi}.
\label{e7}
\end{equation}
This 3D scaling, distinct from the quantized edge conductance of 2D quantum spin Hall systems, can be intuitively understood from the bulk-boundary correspondence of the SOC-gapped phase, which hosts Dirac-cone-like surface states within the bulk gap \cite{SM}.

The quantized scaling law of Eq.~(\ref{e7}) governs both the conventional SHC induced by Rashba-type SOC and the unconventional SHC induced by Weyl-type SOC. When both SOC types coexist, their mixing ratio controls the relative weights of conventional versus unconventional contributions, yet each nonzero component individually retains the linear quantized dependence on $R$ \cite{SM}. 
Full numerical Kubo calculations confirm this universal scaling without approximation: over a wide parameter range, the SHC values (symbols) fall precisely on the analytical lines predicted by Eq.~(\ref{e7}) [Fig.~\ref{p2}(b)]. Convergence with respect to the integration boundary (set by the spatial extent of the SBC) and $k$-mesh density is carefully verified in \cite{SM}.

\paragraph{SBC distribution.} To uncover the microscopic origin of the quantized SHC, we examine the momentum-space SBC. For the gapped nodal-ring model, two representative components associated with the dominant SHC contributions,
$\Omega_{zy}^{x,\mathrm{Rashba}}$ and
$\Omega_{zx}^{x,\mathrm{Weyl}}$, take the closed forms
\begin{align}
\Omega_{zy}^{x, \text{Rashba}}(\mathbf{k})&= \frac{(-k_x^2+k_y^2+R^2) D_0 v_z \lambda_{\text{so}}}{((k_\rho^2-R^2)^2 D_0^2+k_z^2 v_z^2+k_\rho^2 \lambda_{\text{so}}^2)^{3/2}}, \\
\Omega_{zx}^{x, \text{Weyl}}(\mathbf{k})&=\frac{(-k_x^2+k_y^2-R^2) D_0 v_z \lambda_{\text{so}}}{((k_\rho^2-R^2)^2 D_0^2+k_z^2 v_z^2+k_\rho^2 \lambda_{\text{so}}^2)^{3/2}},
\end{align}
with $k_\rho^2=k_x^2+k_y^2$.
On the $k_z=0$ plane, the SBC is sharply peaked around the original nodal ring defined by $k_\rho = R$ [Figs.~\ref{p2}(c,d)], identifying it as the main spin-transport hotspot. Consistent with the $x$-polarized SHC component, the SBC intensity follows the local $S_x$ projection, becoming maximal where the spin texture aligns with $S_x$ and diminishing as it rotates away.

To reveal the geometric origin of the response, we transform to cylindrical coordinates $(r,\theta,k_z)$ and decompose the SHC azimuthally. For fixed $\theta$, integration of the SBC over $(r,k_z)$ (the green line in Fig.~\ref{p2}(c)) yields remarkably simple closed-form $\theta$-resolved expressions:
\begin{align}
\sigma^{x,\mathrm{R}}_{zy}(\theta) &= \frac{e}{4\pi}\,R\,\sin^2\theta, \label{eq:wedge_R}\\
\sigma^{x,\mathrm{W}}_{zx}(\theta) &= -\frac{e}{4\pi}\,R\,\cos^2\theta. \label{eq:wedge_W}
\end{align}
These angular dependences reflect the underlying spin texture. In the Rashba case, for example, at $\theta=\pi/2$ and $3\pi/2$, the local spin texture is fully projected onto $S_x$, making $S_x$ effectively a good quantum number and yielding one half of a spin Hall quantum on each of these two $\theta$-resolved cuts. By contrast, at $\theta=0$ and $\pi$, the vanishing $S_x$ projection eliminates the contribution to $\sigma^{x}_{zy}$. Integrating $\sigma(\theta)$ over the azimuthal angle recovers the total quantized SHC, with the plateau fixed by the nodal-ring geometry $R$. 
The absence of \(D_0\) and \(v_z\) in the \(\theta\)-resolved SHC indicates that anisotropic band dispersion merely reshapes the local SBC hotspot, leaving the integrated geometric response unchanged. Detailed derivations, symmetry constraints, and their connection to the Rashba and Weyl spin textures, together with the effect of band tilting and dispersion, are provided in \cite{SM}.

\begin{figure}[b]
    \centering
    \includegraphics[width=0.48\textwidth]{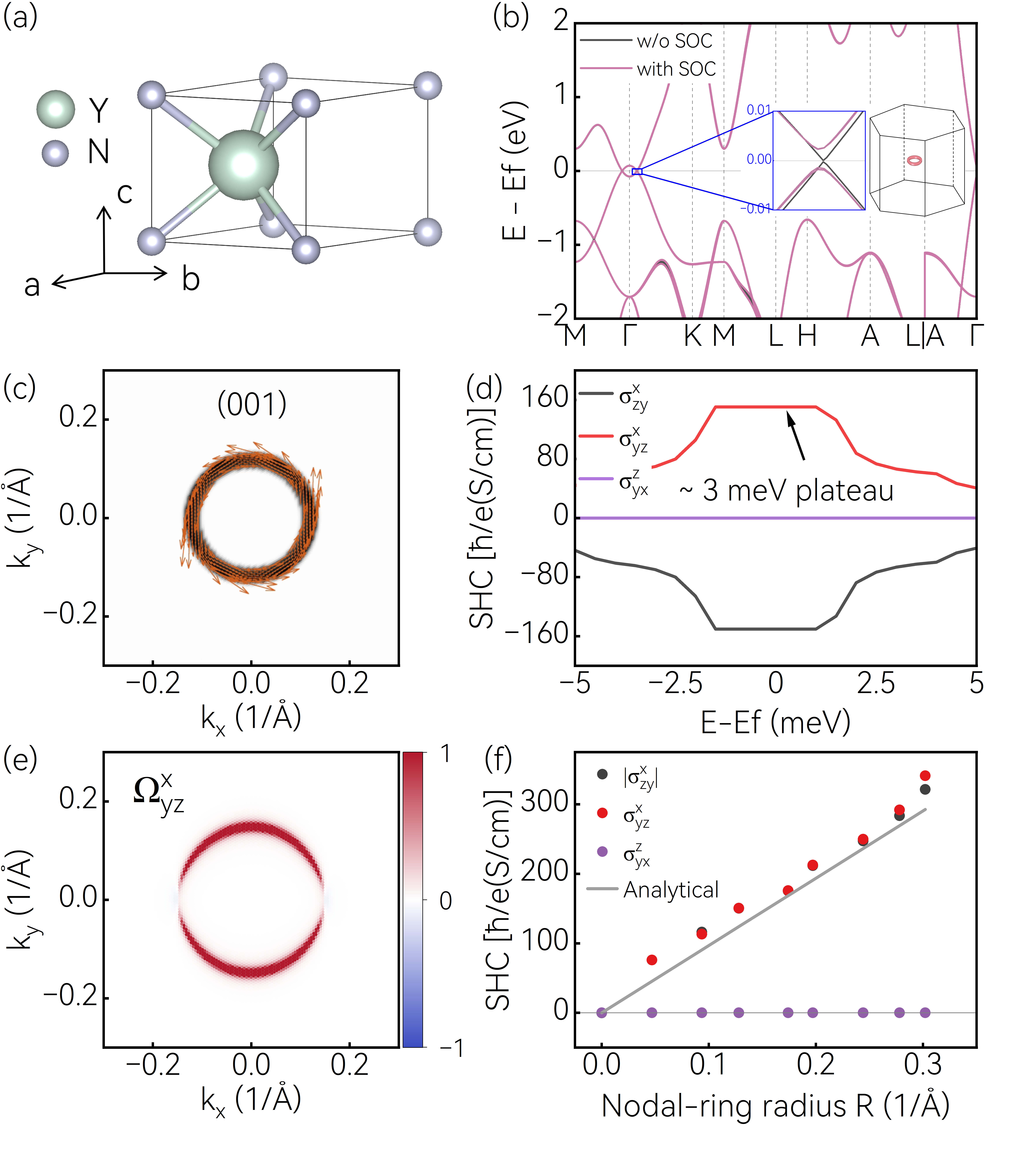}
    \caption{(a) Primitive unit cell of YN, with green and gray spheres denoting Y and N atoms. (b) DFT band structure without (black line) and with SOC (pink line); the inset shows the three-dimensional nodal ring in the \(k_z=0\) plane. The Fermi level is set to zero. (c) Surface-state spin texture on the (001) plane with SOC; arrows denote the in-plane spin orientation, exhibiting pure Rashba-type spin-momentum locking. (d) Energy-dependent representative symmetry-allowed conventional SHC components; symmetry-equivalent components are omitted for clarity. (e) SBC distribution of $\Omega_{yz}^{x}$ on the \(k_z=0\) plane, with red and blue colors denoting opposite signs; the color bar is in arbitrary units. (f) SHC versus nodal-ring radius \(R\) under uniaxial strain; symbols denote DFT results and the solid gray line represents the analytical scaling relation, $\sigma \approx 968.5\,R\,\big[\hbar/e (\mathrm{S}/\mathrm{cm})\text{\AA}\big]$. 
   }\label{p3} 
\end{figure}

\paragraph{Material Realization.} To identify candidate materials, we first establish the symmetry requirements. A flat nodal ring requires a mirror symmetry (e.g., $M_{001}$) to pin it to a specific plane \cite{fangTopologicalNodalLine2015,wengTopologicalNodelineSemimetal2015a}, while broken inversion symmetry allows an SOC-induced spin texture. Furthermore, to realize Weyl-type SOC and the associated unconventional SHC, the mirror plane containing the nodal ring must lack any additional in-plane mirror symmetry \cite{songCoexistenceLargeConventional2020}, which would otherwise suppress the radial spin texture. The allowed magnetic point groups that satisfy these conditions are listed in Table~V of \cite{SM}.

Among the materials satisfying these criteria, we focus on YN, which crystallizes in the noncentrosymmetric hexagonal space group ${P\overline{6}m2}$ [Fig.~\ref{p3}(a)]. Density functional theory (DFT) calculations \cite{huangTunableTopologicalSemimetal2018,kresseEfficientIterativeSchemes1996} without SOC show a single nearly perfectly flat nodal ring at the Fermi level in the \(k_z=0\) plane, enforced by the mirror symmetry $M_{001}$ [Fig.~\ref{p3}(b)]. The topological character of the nodal ring is verified by a \(\pi\) Berry phase for loops encircling the ring \cite{asbothShortCourseTopological2016,fangBulkTopologicalInvariants2012,soluyanovComputingTopologicalInvariants2011a,yuEquivalentExpression22011} and by drumhead surface states \cite{wangDiracSemimetalTopological2012a,ryuTopologicalOriginZeroEnergy2002,wanTopologicalSemimetalFermiarc2011}; see \cite{SM}.


\paragraph{SHC of YN.} When SOC is included, a small global gap ($\sim$3 meV) opens along the nodal ring [Fig.~\ref{p3}(b)], driving YN into a 3D topological insulator phase with $Z_2$ indices (1;000) \cite{fuTopologicalInsulatorsThree2007,fuTopologicalInsulatorsInversion2007,poSymmetrybasedIndicatorsBand2017,songQuantitativeMappingsSymmetry2018,tangComprehensiveSearchTopological2019}. The resulting topology is evidenced by a Dirac-like surface state bridging the bulk valence and conduction bands and exhibiting Rashba spin-momentum locking [Fig.~\ref{p3}(c)], consistent with the minimal nodal-ring model. The presence of a single flat nodal ring at the Fermi level, together with its intrinsic Rashba‑type SOC, makes YN an ideal platform for testing our theoretical predictions.

We compute the full SHC tensor using maximally localized Wannier functions (MLWFs) \cite{qiaoCalculationIntrinsicSpin2018,marzariWannier90MaximallyLocalized2012,MOSTOFI2008685}, with the resulting tight-binding Hamiltonian verified to faithfully preserve the material symmetry required for accurate SHC evaluation \cite{SM}. Symmetry constrains the SHC tensor of YN to three independent antisymmetric pairs \cite{kleinerSpaceTimeSymmetryTransport1966,seemannSymmetryimposedShapeLinear2015}, $\sigma^x_{yz}=-\sigma^y_{xz}$, $\sigma^y_{zx}=-\sigma^x_{zy}$, and $\sigma^z_{xy}=-\sigma^z_{yx}$, as confirmed by numerical calculations shown in Fig.~\ref{p3}(d). This tensor structure also directly reflects the underlying Rashba spin texture: the dominant in-plane spin polarization ($S_x \approx S_y \gg S_z$) yields sizable and comparable in-plane SHC components, while the out-of-plane polarization and the corresponding $\sigma^z$ response are negligible. To uncover the microscopic origin of this response, we examine the SBC of the representative component
$\sigma^{x}_{yz}$ on the $k_z=0$ plane [Fig.~\ref{p3}(e)]. Its sharp localization around the gapped nodal ring confirms the SOC-induced band inversion there as the dominant source of the large, quantized SHC.

To test the universality of the predicted scaling, we express the quantized plateau $(e^2/h) \cdot (\hbar/2 e) \cdot (\pi R/2\pi)$ in practical units as $\sigma \approx 968.5\,R\,\big[\hbar/e (\mathrm{S}/\mathrm{cm})\text{\AA}\big]$ (with $R$ in $\text{\AA}^{-1}$). We then perform a direct DFT-based computational experiment on YN. Applying uniaxial strain along the $c$ axis continuously tunes the nodal-ring radius $R$ within the nodal-ring regime, while preserving the topological character of the band structure. As shown in Fig.~\ref{p3}(f), the DFT SHC values (points) fall on the straight line predicted by our minimal model (solid gray line), confirming the linear scaling $\sigma_{\alpha \beta}^{S, 3D}=\sigma_0^{S,2D} \cdot (\pi R/2 \pi)$. This agreement is notable because the real electronic structure of YN contains orbital complexities absent in the toy model \cite{SM}, demonstrating that the quantized spin Hall response is topologically robust and essentially model‑independent.

We further show that the SOC symmetry can be engineered to tailor the SHC tensor: starting from a $\sqrt{3}\times\sqrt{3}\times1$ supercell, we break $M_{110}$ by a small in-plane rotation of each Y atom about the underlying N sites, while preserving $C_{3[001]}$ and $M_{001}$. This distortion mixes Rashba and Weyl SOC, inducing unconventional SHC components alongside conventional ones. Crucially, each component individually retains the linear proportionality to $R$, with their relative amplitudes determined by the Rashba-Weyl mixing ratio, as predicted by our generalized model \cite{SM}. Thus, strain and symmetry provide independent knobs to control, respectively, the magnitude and the tensor symmetry of the quantized SHC.

\paragraph{Conclusion and perspective.} We have demonstrated that the momentum-space geometry of a topological nodal ring can quantize the 3D SHC. The universal scaling law $\sigma_{\alpha \beta}^{S, 3D}=\sigma_0^{S,2D} \cdot (\pi R/2 \pi)$ arises from the integrated quantized slice contributions, establishing the nodal radius $R$ as a direct geometric knob for the SHC amplitude. Crucially, the SOC symmetry acts as a second, independent switch: Rashba-type SOC generates a purely conventional SHC tensor, whereas Weyl-type SOC activates unconventional components. 
DFT calculations in symmetry-selected YN validate both the scaling and tensor selectivity, with strain tuning $R$ and the SHC magnitude, while symmetry breaking tailors the SOC character. 
The same geometric tunability is expected to appear in experimentally relevant spin Hall resistivity signals, although tensor inversion and longitudinal transport channels can modify the strict linearity in $R$ \cite{SM}. Multiterminal quantum-transport calculations based on the Landauer--B\"uttiker formalism \cite{Buttiker1986,Datta1995,Li2005,Chu2011,Zhang2012} show the same $R$-dependent trend, supporting the observability of the intrinsic SHC scaling \cite{SM}. Beyond the present SHC response, nodal-ring geometry can influence density of states, optical and magnetic responses, and drumhead surface states, suggesting that it may serve as a broader geometric control parameter for quantum transport and response phenomena. This complements previous studies where the nodal-loop size determines the crossover between 3D weak localization and 2D weak antilocalization in disordered nodal‑line semimetals \cite{chenWeakLocalizationAntilocalization2019}.

This work provides a blueprint for engineering spin transport and topological quantum
responses in three dimensions, with potential extensions to other 3D topological phases,
such as nodal-line and nodal-surface systems. The ability to switch between conventional and unconventional SHC components on demand offers a route to tailoring spin polarization for specific spin-orbit-torque architectures. More broadly, the parallel between SHC scaling with the nodal-ring radius and AHC scaling
with the Weyl-pair separation suggests a unified framework for geometric quantization in
3D topological matter with distinct Hall-type responses, including thermal, orbital,
nonlinear, and higher-order responses, each potentially governed by its own geometric
parameter.

\begin{acknowledgments}
This work is supported by the National Key R\&D Program of China (Grant No. 2022YFA1403500), the NSF of China (Grants Nos. W2511003, 12204037, 12234003, and 12321004), and the Beijing Institute of Technology Research Fund Program for Young Scholars.
\end{acknowledgments}

%

\end{document}